\documentclass[12pt]{article}
\usepackage{epsfig,amssymb,amsmath,psfrag,subfigure,rotate,color,wasysym}

\allowdisplaybreaks
\newcommand{\insertfig}[2]{\includegraphics[width=#1cm]{#2}}

\def\XXint#1#2#3{{\setbox0=\hbox{$#1{#2#3}{\int}$ }
\vcenter{\hbox{$#2#3$ }}\kern-.6\wd0}}

\def \be  {\begin{equation}}
\def \ee  {\end{equation}}
\def \ba  {\begin{eqnarray}}
\def \ea  {\end{eqnarray}}
\def \baa {\begin{eqnarray*}}
\def \eaa {\end{eqnarray*}}
\def \lab #1 {\label{#1}}

\newcommand\re[1]{(\ref{#1})}
\def\d{\hbox{{d}\kern-.20em\hbox{l}}}

\def \matrix #1 {\left(\begin{array}{cc} #1 \end{array}\right)}

\newcommand \vev [1] {\langle{#1}\rangle}

\newcommand{\ft}[2]{{\textstyle\frac{#1}{#2}}}
\numberwithin{equation}{section}
\begin{document}

\begin{titlepage}

\thispagestyle{empty}

\vspace*{3cm}

\centerline{\large \bf Vacuum expectation value of twist fields}
\vspace*{1cm}

\centerline{\sc A.V.~Belitsky}

\vspace{10mm}

\centerline{\it Department of Physics, Arizona State University}
\centerline{\it Tempe, AZ 85287-1504, USA}

\vspace{2cm}

\centerline{\bf Abstract}

\vspace{5mm}

Twist fields emerge in a number of physical applications ranging from entanglement entropy to scattering amplitudes in four-dimensional gauge theories.
In this work, their vacuum expectation values are studied in the path integral framework. By performing a gauge transformation, their correlation functions
are reduced to field theory of matter fields in external Aharonov-Bohm vortices. The resulting functional determinants are then analyzed within the zeta function 
regularization for the spectrum of Bessel zeros and concise formulas are derived.

\end{titlepage}

\setcounter{footnote} 0

\newpage




\section{Introduction}

Twist fields (or operators) $V_\alpha$ are ubiquitous in quantum physics. They emerge in various contexts, ranging from orbifold conformal field 
theories \cite{Dixon:1986qv,Knizhnik:1987xp,Bershadsky:1987jk} to correlation functions defining entanglement entropy 
\cite{Holzhey:1994we,Cardy:2007mb,Blondeau-Fournier:2016rtu} or scattering amplitudes in four-dimensional gauge theories 
\cite{Basso:2014jfa,Belitsky:2015lzw,Bonini:2016knr}, just to name a few. Their defining property is a nontrivial monodromy when an elementary field $X$ 
is taken around it, as it acquires a phase ${\rm e}^{2 \pi i \alpha}$. This corresponds to the short-distance operator product expansion
\begin{align}
\label{TwistOPE}
X (z) V_\alpha (0) = | z |^{- \alpha} :X (0) V_\alpha (0): + \dots
\, .
\end{align}

The $V_\alpha$ operators are reminiscent of the (dis)order variables in the two-dimensional Ising model introduced by Kadanoff and Ceva \cite{Kadanoff:1970kz},
with the well-known Mandelstam fermion \cite{Mandelstam:1975hb} being the product of the two. Their continuum field theory as quantum kinks (with 
$\alpha = 1/2$) was proposed by Sato, Miwa and Jimbo \cite{Sato:1978ht} and Schroer and Truong \cite{Schroer:1978sy}, where they were shown to 
be defined by composite, non-polynomial and, therefore, not manifestly local operators of elementary quantum fields. These considerations were
further generalized in Refs.\ \cite{Marino:1981we,Schroer:1982iq} to theories with matter fields being fundamental scalars (along with fermions) within the 
framework of the path integral. It was demonstrated there that correlation functions of twist fields in two-dimensional quantum field theory can be reformulated 
as statistical mechanics of matter fields with Aharonov-Bohm fluxes, with relativistic invariance of the former stemming from gauge invariance of functional 
determinant in the latter.

However, except for twist operators with elementry fermion fields, where techniques based on bosonization \cite{Dixon:1986qv,Knizhnik:1987xp,Bershadsky:1987jk} 
or Fujikawa anomaly \cite{Fujikawa:1980eg} in the path integral measure \cite{Marino:1981we} allow one to bypass direct calculation of determinants, no explicit 
results are available in this formalism for correlation functions of $V_\alpha$'s for theories with fundamental scalars. In this paper we start filling this gap and 
provide a calculation for the vacuum expectation value of a single twist operator within the framework of the zeta function regularization \cite{Hawking:1976ja}. 

Our subsequent presentation is organized as follows. In the next section, we provide a brief recapitulation of field-theoretical description of the twist operator
in the language of the Aharonov-Bohm vortices and relate its vacuum expectation values to the diagonalization problem of Laplacian in external gauge field.
The problem is shown to be reduced to the analysis of the Bessel zeta function. As a starting point in the explanation of the calculational framework, we demonstrate
all salient features for a simpler example of a double Barnes zeta function in Sect.\ \ref{Section2DHurwitz}. Next, in Sect. \ref{VEVsection}, we apply it to the case at 
hand and provide the main result of this paper and summarize in Sect.\ \ref{DiscSection}.

\section{Twist field as Aharonov-Bohm vortex}

The twist field that obeys the operator algebra \re{TwistOPE} was shown to admits the following form \cite{Marino:1981we,Schroer:1982iq}
\begin{align}
\label{TwistFieldOperator}
V_\alpha (x) = \exp \left( 2 \pi i \alpha \int_{C_{[x,\infty]}} d z_\mu \varepsilon_{\mu\nu} j_\nu (z) \right)
\, ,
\end{align}
where the integral in the exponential runs along an arbitrary contour $C_{[x,\infty]}$ from the operator insertion to infinity. It creates a bare Bloch wall in the
terminology of Ref.\  \cite{tHooft:1977nqb} on the curve $C_{[x,\infty]}$. The integrand is determined by the the U(1) current, which for the complex scalar
reads
\begin{align}
j_\mu =  \left( \partial_\mu \phi^\ast  \right) \phi - \phi^\ast \left( \partial_\mu \phi \right)
\, .
\end{align}
The vacuum expectation value of the (s)calar twist operator is then determined by the path integral
\begin{align}
\vev{V^{\rm s}_\alpha}
=
\int [D \phi^\ast] [D \phi] {\rm e}^{- \int d^2 z \left( D_\mu \phi (z) \right)^\ast D_\mu \phi (z) }
\left/
\int [D \phi^\ast] [D \phi] {\rm e}^{- \int d^2 z \left( \partial_\mu \phi (z) \right)^\ast \partial_\mu \phi (z) }
\right.
\, ,
\end{align}
where the exponent of the twist operator was cast in the form of a two-dimensional integral of the interaction of the U(1) scalar current with an external gauge field
\begin{align}
\label{GaugeField}
2 \pi \int_{C_{[x,\infty]}} d z_\mu \varepsilon_{\mu\nu} j_\nu (z) = \int d^2 z j_\mu (z) A_\mu (z;x)
\, , \qquad
A_\mu (z;x) = - 2 \pi \varepsilon_{\mu\nu} \int_{C_{[x, \infty]}} d z^\prime_\nu \delta^{(2)} (z^\prime - z)
\, ,
\end{align}
and, in addition, a quadratic term in $A_\mu$ was added in order to eliminate the path dependence of the correlation function \cite{Marino:1981we}. In this manner, 
the exponential weight becomes a conventional scalar action with the minimal U(1) coupling via the covariant derivative
\begin{align}
\label{CovariantDerivative}
D_\mu = \partial_\mu - i \alpha A_\mu
\, .
\end{align}
In what follows, we will set $x=0$ and drop this argument altogether. In the language of gauge fields, the independence of the choice of $C$ gets reformulated as 
a gauge independence of the functional integral under a phase transformation of the scalar field. We will use this property momentarily. The resulting path integral 
can be computed as
\begin{align}
\vev{V^{\rm s}_\alpha} = \frac{\det \partial^2}{\det D^2}
\, .
\end{align}
Making use of the freedom to choose the path for the line integral \re{GaugeField}, we can align it with the positive $x$-axis, $d z^\prime_\mu = \delta_{\mu 1} d z^\prime_1$, 
such that its Cartesian components become
\begin{align}
A_1 (z) = 0 \, , \qquad A_2 (z) = 2 \pi \theta (z_1) \delta (z_2)
\, .
\end{align}
The computation of the determinant in the above gauge field requires nontrivial boundary conditions, see, e.g., \cite{Jackiw:1989qp}. One notices, 
however, that this configuration is a two-dimensional analogue of the Dirac string with the magnetic induction $B (z) = \varepsilon_{\mu\nu} \partial_\mu A_\nu (z) = 2 
\pi \delta^{(2)} (z)$. The same field is generated by the Aharonov-Bohm vortex (in polar coordinates)
\begin{align}
A_r (z) = 0 \, , \qquad A_\theta (z) = \frac{1}{r}
\, ,
\end{align}
as can be easily established by conventional analysis, see e.g., \cite{Shnir}, with a regularized potential $ A^{\varepsilon}_\theta (z) = 1/\sqrt{r^2 + \varepsilon^2}$ such that
$B (z) = \lim_{\varepsilon \to 0} \varepsilon^2/[r (r^2 + \varepsilon^2)^{3/2}] = 2 \pi \delta^{(2)} (z)$. The two configurations are in fact related by a gauge transformation 
$A_\mu \to A_\mu + i U^\ast \partial_\mu U$, which cancels the Dirac string and leaves instead the covariant Aharonov-Bohm potential
\begin{align}
A_\mu (z) = \frac{z_\nu \varepsilon_{\nu\mu}}{z^2}
\, .
\end{align}
which defines now the covariant derivative \re{CovariantDerivative}.

To compute the determinant of the covariant Laplacian $D^2$, we have to solve the corresponding spectral problem. To generate a discrete spectrum of eigenvalues, one 
imposes a Dirichlet boundary condition at $r = R$. In polar coordinates, the equation admits the form 
\begin{align}
\left( \frac{\partial}{\partial \ln r} - i \frac{\partial}{\partial \theta} + \alpha \right)
\left( \frac{\partial}{\partial \ln r} + i \frac{\partial}{\partial \theta} - \alpha \right) \Phi (r, \theta) = - E^2 r^2 \Phi (r, \theta)
\, .
\end{align}
Separating the variables
\begin{align}
\Phi (r, \theta) = \mathcal{R} (r) {\rm e}^{i m \theta}
\, ,
\end{align}
with integer $m = 0, \pm 1, \pm 2, \dots$ arising from the single-valuedness of the wave function $\Phi$, the solution for $R (\rho)$ is given in terms of the Bessel function
\begin{align}
\mathcal{R} (r) = J_{|m + \alpha|} (E r) + c N_{|m + \alpha|} (E r)
\, .
\end{align}
The vanishing of $R (r)$ and the origin and the Dirichlet boundary condition $\mathcal{R} (R) = 0$, forces us to set $c = 0$ and provides a quantization condition for the 
eigenenergy $E$,
\begin{align}
E_{m, n}^{(\alpha)} = j_{|m + \alpha|, n}/R
\, , 
\end{align}
where $j_{|m + \alpha|, n}$ are the positive zeros of the Bessel function. We assume that $0 < \alpha < 1$ since, as is well known, only the fractional part of $\alpha$ induces 
a nontrivial  Aharonov-Bohm effect.

The vacuum expectation value of the vortex operator is then given by the ratio of eigenvalue products
\begin{align}
\vev{V^{\rm s}_\alpha} = \prod_{m = - \infty}^\infty \prod_{n=1}^\infty \left( \frac{E_{m,n}^{(\alpha)}}{E_{m,n}^{(0)}} \right)^2
\, .
\end{align}
Making use of the zeta-function regularization \cite{Hawking:1976ja}, we can rewrite this as
\begin{align}
\ln  \frac{\det \partial^2}{\det D^2} 
=
\mathcal{Z}^\prime_1 (0, \alpha) 
+
\mathcal{Z}^\prime_2 (0, \alpha) 
+
\mathcal{Z}^\prime_2 (0, - \alpha)
- \mathcal{Z}^\prime_1 (0, 0) 
- 2 \mathcal{Z}^\prime_2 (0, 0)
\, ,
\end{align}
where the prime stands for the derivative with respect to the first argument and we introduced one- and two-dimensional functions
\begin{align}
\label{Z1Z2}
\mathcal{Z}_1 (s, \alpha) = \sum_{n=1}^\infty \big( E_{0,n}^{(\alpha)} \big)^{- 2 s}
\, , \qquad
\mathcal{Z}_2 (s, \alpha) = \sum_{m=1}^\infty \sum_{n=1}^\infty \big( E_{m,n}^{(\alpha)} \big)^{- 2 s}
\, .
\end{align}
The former, i.e., $\mathcal{Z}_1$, is a generalization of the Riemann zeta function from the spectrum of positive integers to the zeros of the Bessel function.
It was thus dubbed the Bessel zeta function in Ref.\ \cite{BesselZeta,BesselZeta2}, it is also known as the Raleigh function \cite{Raleigh}, for even integers $s = 2,4,6, \dots$.
We will refer to $\mathcal{Z}_{1/2}$ as single/double Bessel zetas.

\section{Warm-up: Barnes double zeta}
\label{Section2DHurwitz}

\begin{figure}[t]
\begin{center}
\mbox{
\begin{picture}(0,100)(200,0)
\put(0,-400){\insertfig{23}{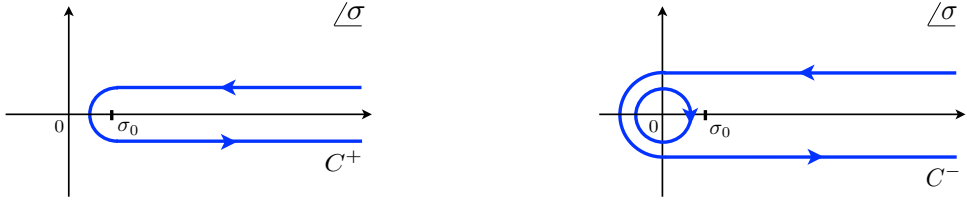}}
\end{picture}
}
\end{center}
\caption{ \label{ContourFig} Integration contour (left panel) for resolvent representation of the zeta function and its deformation (right panel).
}
\end{figure}

Before we move on to the analysis of the Bessel zeta functions, let us consider a simpler example first. For large values of the order of the Bessel function,
the leading two terms in the asymptotic expansion of its zeros read $j_{m + \alpha, n} = \pi (2 n + m + \alpha - 1/2)/2 + O (1/n)$. Thus, it calls for consideration
of a generalization of the Hurwitz zeta function
\begin{align}
\label{HurwitzZeta}
\zeta (s, \alpha) = \sum_{n = 0}^{\infty} (n + \alpha)^{- s}
\end{align}
to the two-dimensional case, i.e., 
\begin{align}
\mathcal{Z} (s, \alpha) = \sum_{m = 1}^{\infty} \sum_{n = 1}^{\infty} (n + m + \alpha)^{- 2 s}
\, ,
\end{align}
which is a particular case of the Barnes double zeta function\footnote{Notice, however, that summations start from one rather than zero which corresponds to the
shift in inhomogeneity $\alpha$.} \cite{BranesZeta}.  In fact, it is rather straightforward to reduce it to the former by means of the Abel-Plana summation formula 
\cite{Erd53}, that it is widely used in analyses of the Casimir effect, see, e.g., \cite{Bordag:2001qi}. One immediately finds Hermite's formula
\begin{align}
\label{2DHurwitzAs1DHurwitz}
\mathcal{Z} (s, \alpha) 
&= - \frac{1}{2} \zeta (2s, 1 + \alpha) + \frac{\zeta (2s - 1, 1 + \alpha)}{2s - 1}
\nonumber\\
&+
i \int_0^\infty \frac{dt}{{\rm e}^{2 \pi t} - 1} \left[  \zeta (2s, 1 + \alpha + i t) - \zeta (2s, 1 + \alpha - i t) \right]
\, .
\end{align}

However, this consideration cannot be extended to the case of the Bessel zeta functions, so one has to resort to other techniques \cite{Spreafico}.
As we will advocate in the next section, it more efficient to deal with the resolvent rather than the heat kernel that is traditionally used in the calculation 
of operator determinants by means of zeta function method \cite{Vassilevich:2003xt}. For the case at hand, we introduce
\begin{align}
\mathcal{R} (\sigma, m + \alpha) = \sum_{n=1}^\infty \frac{1}{\sigma - (n + m + \alpha)^2/(m + \alpha)^2}
\, ,
\end{align}
and write
\begin{align*}
\label{etaResolvent}
\mathcal{Z} (s, \alpha) 
=
\sum_{m=1}^\infty (m + \alpha)^{-2 s} \int_{C^+} \frac{d \sigma}{2 \pi i} \, \sigma^{- s} \, \mathcal{R} (\sigma, m + \alpha) 
\, ,
\end{align*}
where the integration contour $C^+$ is shown in Fig.\ \ref{ContourFig} (left panel) with $\sigma_0 = (2 + \alpha)^2/(1 + \alpha)^2$. Introducing a potential $\mathcal{U}$ 
for the resolvent,
\begin{align}
\mathcal{R} (\sigma, m + \alpha) = - \mathcal{U}^\prime (\sigma, m + \alpha)
\end{align}
we can remove the $\sigma$-derivative off $\mathcal{U}$ by integrating by parts in Eq.\ \re{etaResolvent}, on the one hand, 
\begin{align}
\label{etaPotential}
\mathcal{Z} (s, \alpha) 
= - s
\sum_{m=1}^\infty (m + \alpha)^{-2 s} \int_{C^+} \frac{d \sigma}{2 \pi i} \, \sigma^{- s - 1} \, \mathcal{U} (\sigma, m + \alpha) 
\, ,
\end{align}
and sum-up the infinite series, on the other, making use of the infinite product representation for the Euler gamma functions (or rather their product),
\begin{align}
\mathcal{U} (\sigma, m + \alpha) 
&
= - \ln \prod_{n=1}^\infty \left( 1 - \frac{(m + \alpha)^2 \sigma}{(n + m + \alpha)^2} \right)
\nonumber\\
&
= 
\ln \frac{\Gamma \big(1 + (m + \alpha) (1 - \sqrt{\sigma}) \big) \Gamma \big(1 + (m + \alpha) (1 + \sqrt{\sigma}) \big)}{\Gamma^2 (1 + m + \alpha)}
\, .
\end{align}

The function $\sum_{n = 1}^\infty (m + \alpha)^{-2s} \mathcal{R} (\sigma, m + \alpha)$ is not analytic in $s$ in the vicinity of $s=0$, that is needed for the computation of
the derivative. However, the singularity can only arise from the $1/(m + \alpha)$ term in the large $m$ expansion of $\mathcal{R} (\sigma, m + \alpha)$, as can be 
seen from the Laurent series for the Hurwitz zeta function:
\begin{align*}
\zeta (2 s + 1, 1 + \alpha) = \frac{1}{2 s} - \psi (1 + \alpha) + O (2s)
\, ,
\end{align*}
which will thus reduce the overall power of $s$ from two to one. Thus it is instructive to split $\mathcal{U}$ into two contributions
\begin{align}
\label{UsplitVW}
\mathcal{U} (\sigma, m + \alpha) = \mathcal{V} (\sigma, m + \alpha) + \frac{1}{m + \alpha} \mathcal{W} (\sigma)
\, ,
\end{align}
where $\mathcal{W} (\sigma)$ can immediately be found from the asymptotic Stirling series for the Euler gamma functions in the left-hand side by going to
the first power suppressed term \cite{Erd53}. It is 
\begin{align}
\mathcal{W} (\sigma) = \frac{1}{6 (1 - \sigma)}
\, ,
\end{align}
and we find
\begin{align}
\label{etaResolvent}
\mathcal{Z} (s, \alpha) 
=
- s \sum_{m=1}^\infty (m + \alpha)^{-2 s} \int_{C^+} \frac{d \sigma}{2 \pi i} \, \sigma^{- s - 1} \, \mathcal{V} (\sigma, m + \alpha) 
+
\frac{s}{6} \zeta (2s + 1, 1 + \alpha)
\, .
\end{align}

Since for the calculation of the determinant, all we need is to track of the terms linear in $s$, the dominant contribution emerges from the small
and large $\sigma$ regions of the integrand. To have a better convergence in the latter domain, we use the Schwinger representation for the $\sigma^{- s}$ 
factor (with $\Re{\rm e} \, \sigma > 0$) in the integrand
\begin{align}
\label{SchwingerRep}
\sigma^{- s} 
&
= \Gamma^{- 1} (s) \int_0^\infty dt \, t^{s - 1} {\rm e}^{- t \sigma}
\nonumber\\
&
= \Gamma^{- 1} (s) \int_0^1 dt \, t^{s - 1} {\rm e}^{- t \sigma} + O (s^2)
\, ,
\end{align}
with the integral representation in the second line valid up to higher order terms in $s$, which are irrelevant for the current analysis. The function 
$\mathcal{V} (\sigma, m + \alpha)$ is analytic at $\sigma = 0$, we move the integration contour $C^+$ to the left of the imaginary axis and pick up 
the pole of the integrand at the origin,
\begin{align}
\label{ContourShift}
\int_{C^+} \frac{d \sigma}{2 \pi i} \, \frac{{\rm e}^{- t \sigma}}{\sigma} \, \mathcal{V} (\sigma, m + \alpha) 
=
- \mathcal{V} (0, m + \alpha) + \int_{C^-} \frac{d \sigma}{2 \pi i} \, \frac{{\rm e}^{- t \sigma}}{\sigma} \, \mathcal{V} (\sigma, m + \alpha) 
\, ,
\end{align}
as shown in Fig.\ \ref{ContourFig} (right panel), and making use of $\mathcal{V} (0, m + \alpha) =  - \mathcal{W} (0)/(m + \alpha) = - 1/[6 (m + \alpha)]$, as can 
be easily found from Eq.\ \re{UsplitVW}. The remaining integral performed over the contour $C^-$. To evaluate this last term, we rescale the integration variable
$\sigma \to \sigma/t$ and expand the integrand for small $t$. We then can deform $C^-$ contour to large values of $|\sigma|$ and, thus, rely only on the asymptotic 
behavior of $\mathcal{V} (\sigma, m + \alpha)$ \cite{Cheeger84,BruSee85,Spreafico},
\begin{align}
\mathcal{V} (\sigma, m + \alpha) 
&
= \left( m + \alpha + \frac{1}{2} \right) \ln (- \sigma) 
\\
&
+ 
2 \left( m + \alpha + \frac{1}{2} \right) \ln (m + \alpha) + \ln (2 \pi) - 2 \ln \Gamma (1 + m + \alpha) + \dots
\, , \nonumber
\end{align}
with ellipses standing for irrelevant (subleading) contributions. Evaluating emerging $\sigma$ integrals,
\begin{align}
\label{SigmaIntegrals}
\int_{C^-} \frac{d \sigma}{2 \pi i} \, \frac{{\rm e}^{- t \sigma}}{\sigma} \, \ln (- \sigma) = - \ln \left( {\rm e}^{\gamma_{\rm E}} t \right)
\, , \qquad
\int_{C^-} \frac{d \sigma}{2 \pi i} \, \frac{{\rm e}^{- t \sigma}}{\sigma} \, = 1
\, ,
\end{align}
and performing the trivial $t$ integration, we finally obtain\footnote{Had we introduced terms with subleading powers of $1/(m + \alpha)^{n >1}$ in Eq.\ \re{UsplitVW},
these can be immediately seen to induce only $O (s)$ effect in the right hand side of Eq.\ \re{BarnesZeta}. E.g.,  for the next nontrivial term $- (1 + 3 \sigma)/[
(1 - \sigma)^3 (m + \alpha)^3$, we get an extra term in the right-hand side of Eq.\ \re{BarnesZeta} of the form $[1 - (1 + 2 s) \Gamma (2 + s)] \zeta (2s + 3, 1 + \alpha)/180$,
which is obviously $\sim s$ at small $s$. These are thus irrelevant for the calculation of the first derivative of $\mathcal{Z} (s, \alpha)$.}
\begin{align}
\label{BarnesZeta}
\Gamma (s) \mathcal{Z} (s, \alpha) 
&
= \zeta^\prime (2s, 1 + \alpha) + 2  \zeta^\prime (2s - 1, 1 + \alpha) + \frac{1}{6} \Gamma (1 + s) \zeta (2s + 1, 1 + \alpha)
\\
&
- \zeta (2s, 1 + \alpha) \ln (2 \pi)
+
\frac{1}{2}
\left(
\gamma_{\rm E} - \frac{1}{s}
\right)
\left[
\zeta (2s, 1 + \alpha)
+
2 \zeta (2s - 1, 1 + \alpha)
\right]
\nonumber\\
&
+
2 \chi (s, \alpha)
+ O (s)
\, . \nonumber
\end{align}
making use of the definition \re{HurwitzZeta}. Here, we introduced a function
\begin{align}
\label{chiFunction}
\chi (s, \alpha)
=
\sum_{m=1}^\infty (m + \alpha)^{-2 s} \ln \Gamma (1 + m + \alpha) - \frac{1}{12} \zeta (2s + 1, 1 + \alpha)
\, ,
\end{align}
which is finite for $s = 0$, but not each of them separately. We will not perform any further reduction of this result since the current form will be used for 
simplification of the Bessel zeta function, which we are turning to in the following section. In fact, we can find a simple form for the function $\chi$ for $s = 0$,
\begin{align}
\chi (0, \alpha) 
&
= \frac{1}{12} \psi (1 + \alpha) - \frac{1}{2} \left( \frac{1}{2} + \alpha \right) \ln (2 \pi) - \zeta (-1, 1 + \alpha)
\\
&
- \zeta^\prime (-1, 1 + \alpha)
- \frac{1}{2} \zeta^\prime (0, 1 + \alpha) + i \int_0^\infty \frac{dt}{{\rm e}^{2 \pi t} - 1} \ln \frac{\Gamma (1 + \alpha + i t)}{\Gamma (1 + \alpha - i t)}
\, , \nonumber
\end{align}
where we have used the Lerch's formula $\zeta^\prime (0, \alpha) = \ln \Gamma (\alpha)/\sqrt{2 \pi}$.

\section{One-vortex determinant}
\label{VEVsection}

Now we are in a position to address to the main object of the present paper. A key observation for an amenable calculation of the Bessel zeta functions \re{Z1Z2} 
is to deal with the Bessel function per se rather than its zeros. This can be accomplished making use of the infinite product representation \cite{Raleigh}
\begin{align}
\label{BesselProduct}
J_\nu (\sigma) = \frac{\left( \sigma/2 \right)^\nu}{\Gamma (1 + \nu)} \prod_{n=1}^\infty \left( 1 - \frac{\sigma^2}{j_{\nu,n}^2} \right)
\, ,
\end{align}
and then following up the same routine as in the previous section.

\subsection{Single Bessel zeta}

Taking advantage of Eq.\ \re{BesselProduct} allows one to find a concise form of the resolvent. Let us start with $\mathcal{Z}_1$, which reads
\begin{align}
\mathcal{Z}_1 (s, \alpha) = \int_{C^+} \frac{d \sigma}{2 \pi i} \sigma^{-s} \, \mathcal{R}_1 (\sigma, \alpha)
\end{align}
and the potential derived from the resolvent being
\begin{align}
\mathcal{R}_1 (\sigma, \alpha) = - \mathcal{U}^\prime_1 (\sigma, \alpha) 
\, , \qquad
\mathcal{U}_1 (\sigma, \alpha) 
=
-  \ln \frac{I_\alpha (R \sqrt{-\sigma})}{(- \sigma)^{\alpha/2}} 
\, .
\end{align}
Here we passed to the function of imaginary argument for better convergence, $I_\nu (\sigma) = i^{- \nu} J_\nu (i \sigma)$.
As in the previous section, making use of Eq.\ \re{SchwingerRep} and moving the integration contour to the left of the imaginary axis, we find
\begin{align}
\Gamma (s)
\mathcal{Z}_1 (s, \alpha) = \mathcal{U}_1 (0, \alpha)
-
s
\int_0^1 dt \, t^{s - 1} \int_{C^-} \frac{d \sigma}{2 \pi i} \frac{{\rm e}^{- t \sigma}}{\sigma} \, \mathcal{U}_1 (\sigma, \alpha) +  O(s)
\, ,
\end{align}
with
\begin{align}
\mathcal{U}_1 (0, \alpha) = \ln \frac{2^\alpha \Gamma (1 + \alpha)}{R^\alpha}
\, .
\end{align}
The remaining integral is found again by studying the large-$|\sigma|$ asymptotics of the integrand,
\begin{align}
\mathcal{U}_1 (\sigma, \alpha) =  \frac{1}{2} \left( \alpha + \frac{1}{2} \right) \ln (- \sigma) + \frac{1}{2} \ln (2 \pi R) + \dots
\, , 
\end{align}
such that by means of the results \re{SigmaIntegrals}, we finally derive
\begin{align}
\Gamma (s)
\mathcal{Z}_1 (s, \alpha)
=
- \frac{1}{2} \left( \alpha + \frac{1}{2} \right) \left( 2 \ln R - \gamma_{\rm E} + \frac{1}{s} \right) 
+ \ln \frac{2^{\alpha - 1/2} \Gamma (1 + \alpha)}{\sqrt{\pi}} + O (s)
\, ,
\end{align}
with the first derivative being \cite{BesselZeta2,Lesh}
\begin{align}
\mathcal{Z}^\prime_1 (0, \alpha)
=
- \left( \alpha + \frac{1}{2} \right) \ln R + \ln \frac{2^{\alpha - 1/2} \Gamma (1 + \alpha)}{\sqrt{\pi}}
\, .
\end{align}
This is a simple manifestation of the Gelfand-Yaglom theorem \cite{Gelfand} for the determinant of a one-dimensional Laplacian with conical singularity.

\subsection{Double Bessel zeta}

Last but not least, we move on to the two-dimensional Bessel zeta function. We follow to the letter the step-by-step the procedure advocated in 
Sect.~\ref{Section2DHurwitz} such that all symbols used there in generic expressions have to be merely dressed with the label 2. The resolvent is traded 
\begin{align}
\mathcal{R}_2 (\sigma, m + \alpha) = - \mathcal{U}^\prime_2 (\sigma, m + \alpha)
\, ,
\end{align}
for a potential
\begin{align}
\mathcal{U}_2 (\sigma, m + \alpha)
&
=
- \ln \prod_{n=1}^\infty \left( 1 - \frac{(m + \alpha)^2 R^2 \sigma}{j_{m + \alpha, n}^2} \right)
\nonumber\\
&
= - \ln I_{m + \alpha} \left( (m + \alpha) R \sqrt{- \sigma} \right) + (m + \alpha) \ln \frac{(m + \alpha) R \sqrt{- \sigma} }{2}
- \ln \Gamma (1 + m + \alpha)
\, .
\end{align}
Splitting it as in Eq.\ \re{UsplitVW}, the $\mathcal{W}_2$ function accompanying the $(m + \alpha)$-pole is extracted from the uniform expansion of 
the Bessel function \cite{Olver} 
\begin{align}
\mathcal{W}_2 (\sigma) = - U_1 (R \sqrt{- \sigma}) 
\, , \qquad
U_1 (z) = \frac{1}{8} (1 + z^2)^{- 1/2} - \frac{5}{24} (1 + z^2)^{-3/2}
\, ,
\end{align}
with the integral of this being
\begin{align}
\int_{C^+} \frac{d \sigma}{2 \pi i} \sigma^{- s - 1} \, \mathcal{W}_2 (\sigma) 
=
-  \frac{\Gamma (s + \ft12)}{12 \Gamma (s) \Gamma (\ft12)} \left( \frac{1}{s} + 5 \right) R^{2s}
\, . 
\end{align}
The shift of the integration contour, as in Eq.\ \re{ContourShift}, allows one to pick the contribution at $\sigma = 0$ 
\begin{align}
\mathcal{V}_2 (0, m + \alpha) = - \frac{1}{12 (m + \alpha)}
\, ,
\end{align}
and evaluate the rest in the large-$|\sigma|$ asymptotic domain
\begin{align}
\mathcal{V}_2 (\sigma, m + \alpha) = 
&
\frac{1}{2} \left( m + \alpha + \frac{1}{2} \right) \ln (- \sigma)
\nonumber\\
&
+ \left( m + \alpha + \frac{1}{2} \right) \ln \big( (m + \alpha) R \big) - \ln \frac{2^{m + \alpha - 1/2} \Gamma (1 + m + \alpha)}{\sqrt{\pi}}
+ \dots
\, .
\end{align}
The remaining steps are identical to the ones in the previous section and we ultimately find
\begin{align}
\Gamma (s)
\mathcal{Z}_2 (s, \alpha)
&
=
\zeta^\prime (2s - 1, 1 + \alpha) + \frac{1}{2} \zeta^\prime (2s, 1 + \alpha) 
\nonumber\\
&
+
\frac{1}{2} 
\left(
\gamma_{\rm E} - \frac{1}{s} - 2 \ln \frac{R}{2}
\right)
\zeta (2s - 1, 1 + \alpha)
+
\frac{1}{4} 
\left(
\gamma_{\rm E} - \frac{1}{s} - 2 \ln ( 2 \pi R )
\right)
\zeta (2s, 1 + \alpha)
\nonumber\\
&
+
\frac{\Gamma (s + \ft12)}{12 \Gamma (\ft12)} (1 + 5 s) \zeta (2s + 1, 1 + \alpha) R^{2s} 
+ 
\chi (s, \alpha) + O (s)
\, .
\end{align}
Notice the appearance of the very same function \re{chiFunction} which allows us to eliminate it in favor of a concise representation for the 
Barnes double zeta function \re{2DHurwitzAs1DHurwitz}. This concludes all the necessary calculations required for evaluation of 
the determinant.

\section{Discussion}
\label{DiscSection}

Calculation of the first derivative at $s=0$ of the deduced one- and two-dimensional Bessel zeta functions provides the result for the vacuum expectation value 
of the scalar twist operator,
\begin{align}
\vev{V^{\rm s}_\alpha} = c^{\rm s}_\alpha R^{- h^{\rm s}_\alpha}
\, ,
\end{align}
where $h^{\rm s}_\alpha = \alpha (1 - \alpha)$ is the well-known conformal dimension of the twist field \cite{Dixon:1986qv} and the normalization constant is
\begin{align}
c^{\rm s}_\alpha
&
=
2^{\alpha (1 - \alpha)} \Gamma (1 + \alpha) \exp \Big( \chi (0,\alpha) + \chi (0, - \alpha) -2 \chi (0,0)  - \ft{1}{12} [\psi (1 - \alpha) + \psi (1 + \alpha) + 2 \gamma_{\rm E}] \Big) 
\, .
\end{align}
In this derivation, we reduced $\zeta (- n, \alpha)$ for negative integer values $s = - n$ to the Bernoulli polynomials $\zeta (- n, \alpha) = - B_{n+1} (\alpha)/(n + 1)$ with 
the first two being $B_1 (\alpha) = \alpha - \ft12$, $B_2 (\alpha) = \alpha^2 - \alpha + \ft16$, see, e.g., \cite{Erd53}.

In a similar fashion, one can compute the one-point function of the fermion twist operator by inserting the fermion U(1) current $j_\mu = \bar\psi \sigma_\mu \psi$
in Eq.\ \re{TwistFieldOperator}. This time the determinant of the Dirac operator has to be evaluated with spectral boundary conditions as was advocated in Ref.\
\cite{Marino:1981we} and it reads
\begin{align}
\vev{V^{\rm f}_\alpha} 
= \frac{\det {\not\!\!D}}{\det {\not\!\partial}}
=
- \mathcal{Z}^\prime_2 (0, \alpha) 
-
\mathcal{Z}^\prime_2 (0, - \alpha)
+ 2 \mathcal{Z}^\prime_2 (0, 0)
=
c^{\rm f}_\alpha R^{- h^{\rm f}_\alpha}
\, ,
\end{align}
with the conformal dimension $h^{\rm f}_\alpha = \alpha^2$ \cite{Dixon:1986qv,Knizhnik:1987xp,Bershadsky:1987jk} and the normalization being
\begin{align}
c^{\rm f}_\alpha
&
=
2^{\alpha^2} \exp \Big( - \chi (0,\alpha) - \chi (0, - \alpha) + 2 \chi (0,0)  + \ft{1}{12} [\psi (1 - \alpha) + \psi (1 + \alpha) + 2 \gamma_{\rm E}] \Big) 
\, .
\end{align}

The technique presented in the main body of the paper can be used as a stepping stone for consideration of higher point functions, in particular, it is feasible
to apply it to two point correlations. However, for three points and beyond, a more efficient framework would be to use the Burghelea-Friedlander-Kappeler gluing 
formula \cite{BFK92} for determinants on surfaces with conical singularities, as was recently discussed in Ref.\ \cite{Sher15}. Here, the vacuum expectation
value found above becomes an intrinsic building block.

\vspace{1cm}

This research was supported by the U.S. National Science Foundation under the grant PHY-1403891.


\end{document}